\documentclass[11pt]{article}
\begin{document}

\title{{\bf Supersymmetric Polytropic Gas Dynamics}}
\author{Ashok Das\\
Department of Physics and Astronomy,\\
University of Rochester,\\
Rochester, NY 14627-0171\\
USA\\
\\
and\\
\\
Ziemowit Popowicz \\
Institute of Theoretical Physics, \\
University of Wroc\l aw,\\
50-205 Wroclaw\\ 
Poland.}
\date{}
\maketitle

\begin{center}
{ \bf Abstract}
\end{center}

We construct the $N=1$ supersymmetric extension of the polytropic gas
dynamics. We give both the Lagrangian as well as the Hamiltonian
description of this system. We construct the infinite set of \lq\lq
Eulerian'' conserved charges associated with this system and show that
they are in involution, thereby proving complete integrability of this
system. We construct the SUSY -B extension of this system as well and
comment on the similarities and differences between this system and an
earlier construction of the supersymmetric Chaplygin gas. We also
derive the $N=1$ supersymmetric extension of the elastic medium
equations, which, however, do not appear to be integrable.

\newpage

\section{Introduction:}

Hydrodynamic systems are dispersionless systems, which have been
studied from a variety of points of view \cite{1}-\cite{5}. In $1+1$
dimensions,  they
encompass a large class of physical systems such as the
polytropic gas \cite{6}, the elastic medium equations \cite{6}, Born-Infeld
equations \cite{7,8},\dots , all of which are integrable. Some of these
dispersionless systems \cite{9} can be related to string theory, membrane
theory and topological field \cite{10} theories.

Although there has been an increased interest in studying such systems in
recent years, a lot remains to be understood about such systems. For
example, it is only recently that a non-standard Lax description
for the polytropic gas \cite{11} (and the Born-Infeld equations
\cite{12})  has been found,
which describes various properties of the system in a unified
manner. A non-standard Lax representation, however, is not useful for
various generalizations and a standard Lax representation has been
obtained only now which brings in the interesting connection between
such systems and the Lucas and Fibonacci polynomials \cite{13}. Such a
representation will, no doubt, be very useful.

Considering the intimate connection between such systems and
string/membrane/topological theories, it is surprising that a
systematic study of the supersymmetrization of such systems does not
exist so far. It is, of course, well known that supersymmetrization of
integrable systems is a notoriously difficult problem. For example, a
complete classification of all possible supersymmetrizations of the
simplest of the integrable systems - the KdV equation - is still
lacking. On top of that, in the case of supersymmetric dispersionless
systems,  a systematic procedure for constructing the Lax description
remains an open question (the Lax description has so far been
constructed only for three systems \cite{14,15} and they have a much more
complicated structure than their dispersive counterparts). However,
a systematic study of the supersymmetrization of such hydrodynamic
systems  is an
important question and, in this letter, we take a first step in this
direction by constructing the $N=1$ supersymmetric polytropic gas dynamics.

It is important to point out here that there have been some earlier
attempts \cite{16,17} at specific supersymmetrizations of such systems. For
example, in \cite{17}, a supersymmetrization of the  Chaplygin gas (which
corresponds to the specific choice of $\gamma =-1$ in the polytropic
gas) has been obtained  from the superstring theory in $3$-dimensions
with a  $kappa$ supersymmetry after gauge fixing. This
supersymmetric system of equations has the peculiar feature that the
two bosonic equations do not have any fermion terms. Within the
context of integrable systems, such supersymmetrizations are well
understood and, in fact, in the early literature on the subject, such
supersymmetrizations were discarded as being \lq\lq trivial''
\cite{18}. It  is
only after their connection with superstrings (through matrix models)
was established that they have been studied vigorously and are called
SUSY -B extensions \cite{19}. Since the system in \cite{17} is
obtained from  a string
theory, it is not surprising that it is quite analogous to a -~B
extension of the original bosonic system of equations. However, as we
will  point
out later, it is not exactly a conventional  -~B extension either.

Since we do not yet know how to systematically construct a Lax
representation for supersymmetric dispersionless systems, we have
supersymmetrized the system directly at the level of the equations of
motion as well as the Hamiltonian structure. In section {\bf 2}, we
will review some of the essential features of the bosonic polytropic
gas dynamics that are useful for our discussion of the SUSY system. In
section {\bf 3},
we construct the supersymmetric system of equations which depends on a
free parameter, as is normally the case in the supersymmetrization of
integrable systems. We show that this system is Hamiltonian only for a
particular value of the parameter. We construct the infinite set of
conserved quantities associated with this system. Surprisingly, the
second set of infinite conserved charges (that occurs in the bosonic
system) reduces to only one local conserved charge. We bring out
and comment on various other interesting features of this system, such
as the second Hamiltonian structure, and show that the
conserved charges are in involution, thereby proving complete
integrability of the system. In section {\bf 4}, we present the
-~B extension of the polytropic gas dynamics and compare this with the
model presented in \cite{17}. We end with a brief conclusion in section {\bf
5}. All our calculations have also been checked using the symbolic
program REDUCE \cite{19'} with the special supersymmetry package
\cite{19''}. 

\section{Polytropic gas dynamics:}

The equations describing the polytropic gas dynamics in 1+1 dimensions
are \cite{6}
\begin{eqnarray}
\frac{\partial v}{\partial t} & = & ( vu )_{x}\nonumber\\
\frac{\partial u}{\partial t} & =  & \left( {u^{2}\over 2} + {v^{\gamma
-1}\over (\gamma
-1)}\right)_{x},\qquad \gamma\neq 0,1\label{1}
\end{eqnarray}
where $u$ and $v$ are the two dynamical variables and
they belong to the class of equations known  as  equations of
hydrodynamic type. The system of polytropic gas equations include a
large class of physical systems corresponding to different values of
the exponent $\gamma$. In particular, for  $\gamma=-1$, they describe
the Chaplygin gas, or the Born - Infeld equation. The dispersionless
limit of the two boson equations corresponds to $\gamma = 2$.
For the case $\gamma = 3$, under the simple change of variables 
\begin{equation}
A = u + v, \qquad B = u - v 
\end{equation}
the equations reduce to
\begin{equation}
\frac{\partial A}{\partial t} = ( A^2)_{x}, \qquad 
\frac{\partial B}{\partial t} = ( B^2 )_{x}\label{1'}
\end{equation}
This system of equation can be though as the dispersionless limit of
two  noninteracting Korteweg de Vries  equations (also known as Riemann
equations). The case $\gamma = 4$
can be  thought of as the dispersionless limit of the 
Boussinesq hierarchy. Let us note that, in general, the
fields $u$ and $v$ can be assigned the canonical dimensions $[v] = 2 $
and  $[u] = \gamma -1$. 

The polytropic gas admits two infinite sets of conserved charges
\cite{5,11}.  The
first set, also called the \lq\lq Eulerian'' conserved charges
\cite{20},  have the explicit form

\begin{equation}
H_{n}^{(1)} = \int dx\,h_{n}^{(1)} = \int dx\,\sum_{k=0}^{[{n\over 2}]}
\,c(k,n)\,u^{n-2k} v^{k(\gamma -1)+1},\qquad n = 0,1,2,\cdots\label{2}
\end{equation}
where $[{n\over 2}]$ stands for the integer part of the fraction, and
\begin{equation}
c(k,n) = {1\over k! (n-2k)!}\,{1\over (\gamma
-1)^{k}}\left(\prod_{l=0}^{k} {1\over l(\gamma -1) + 1}\right)\label{3}
\end{equation}
Explicitly, the first four conserved charges of this infinite set have
the forms
\begin{eqnarray}
H_{0}^{(1)} & = & \int dx\,h_{0}^{(1)} = \int dx\,v\nonumber\\
H_{1}^{(1)} & = & \int dx\,h_{1}^{(1)} = \int dx\, uv\nonumber\\
H_{2}^{(1)} & = & \int dx\,h_{2}^{(1)} = \int dx\, \left({1\over 2!} u^{2}v
+ {v^{\gamma}\over \gamma (\gamma -1)}\right)\nonumber\\
H_{3}^{(1)} & = & \int dx\,h_{3}^{(1)} = \int dx\, \left({1\over 3!} u^{3}v
+ {uv^{\gamma}\over \gamma (\gamma -1)}\right)
\end{eqnarray}

The second infinite set of conserved charges, also known as the
\lq\lq Lagrangian'' conserved charges \cite{20}, have the following
closed  forms
\begin{equation}
H_{n}^{(2)} = \int dx\,h_{n}^{(2)} = \int dx\,\sum_{k=0}^{[{n\over
2}]}\,c(k,n) u^{n-2k+1} v^{k(\gamma -1)}\label{4}
\end{equation}
The first three of these conserved charges have the explicit forms
\begin{eqnarray}
H_{0}^{(2)} & = & \int dx\,h_{0}^{(2)} = \int dx\, u\nonumber\\
H_{1}^{(2)} & = & \int dx\,h_{1}^{(2)} = \int dx\,\left({1\over 2!}
u^{2} + {v^{(\gamma -1)}\over (\gamma -1)(\gamma -2)}\right)\nonumber\\
H_{2}^{(2)} & = & \int dx\,h_{2}^{(2)} = \int dx\,\left({1\over 3!}
u^{3} + {uv^{(\gamma -1)}\over (\gamma -1)(\gamma -2)}\right)
\end{eqnarray}

The system of polytropic gas equations is Hamiltonian with respect to
three distinct Hamiltonian structures (the operators act on a delta
function), 
\[
{\cal D}_{1} = \left( \begin{array}{cc}
     0 & \partial \\ \partial & 0  
\end{array} \right ),
\]
\[
{\cal D}_{2} =  \left( \begin{array}{cc}
 \partial v^{\gamma -2} + v^{\gamma -2}\partial &  \partial u +
 (\gamma -2)u \partial  \\
 (\gamma -2)\partial u + u \partial & \partial v +
 v \partial 
\end{array} \right )
\] 
\begin{equation}
{\cal D}_{3} = \left( \begin{array}{cc}
\partial uv^{\gamma -2} + uv^{\gamma -2} \partial  
&
\begin{array}{c}
 \partial \Big [ \frac{1}{2} u^2 +
\frac{1}{\gamma-1}v^{\gamma-1}) \Big ] \\ 
+ \Big [ \frac{(\gamma -2)}{2}u^2 + \frac{1}{\gamma-1}v^{\gamma-1}
\Big ] \partial 
\end{array}
 \\
\begin{array}{c}
\partial \Big [ \frac{(\gamma -2)}{2}u^2
+\frac{1}{\gamma-1}v^{\gamma-1} \Big ] \\
+\Big [ \frac{1}{2}u^2 + \frac{1}{\gamma-1}v^{\gamma-1} \Big
]\partial  
\end{array}
& 
\partial uv + 
uv \partial
\end{array} \right )\label{5}
\end{equation}
so that we can write the polytropic gas equations as
\begin{equation}
\left(\begin{array}{c}
u_{t}\\
v_{t}
\end{array}\right) = {\cal D}_{1}\left(\begin{array}{c}
{\delta H_{2}^{(1)}\over \delta u}\\
{\delta H_{2}^{(1)}\over \delta v}
\end{array}\right) = {1\over \gamma}\, {\cal D}_{2}\left(\begin{array}{c}
{\delta H_{1}^{(1)}\over \delta u}\\
{\delta H_{1}^{(1)}\over \delta v}
\end{array}\right) = {\cal D}_{3}\left(\begin{array}{c}
{\delta H_{0}^{(1)}\over \delta u}\\
{\delta H_{0}^{(1)}\over \delta v}
\end{array}\right)
\end{equation}
Similarly, the other \lq\lq Eulerian'' conserved quantities, $H_{n}^{(1)}$,
give the higher order equations of the polytropic gas hierarchy with
these three Hamiltonian structures.

On the other hand, the second set of conserved quantities, with the
Hamiltonian structures of Eq. (\ref{5}), lead to the hierarchy of the elastic
medium equations. For example, the elastic medium equations
\begin{equation}
v_{t} = u_{x},\qquad u_{t} = v^{\gamma -3} v_{x}\label{6}
\end{equation}
can be easily seen to result from
\begin{equation}
\left(\begin{array}{c}
u_{t}\\
v_{t}
\end{array}\right) = {\cal D}_{1}\left(\begin{array}{c}
{\delta H_{1}^{(2)}\over \delta u}\\
{\delta H_{1}^{(2)}\over \delta v}
\end{array}\right) = {1\over \gamma -2}\, {\cal D}_{2}\left(\begin{array}{c}
{\delta H_{0}^{(2)}\over \delta u}\\
{\delta H_{0}^{(2)}\over \delta v}
\end{array}\right)
\end{equation}

The three Hamiltonian structures in Eq. (\ref{5}) are
compatible,  which can
be easily seen from the fact that under the shift $ u \rightarrow 
u+ \lambda $, where $\lambda $ is an arbitrary constant, $ {\cal D}_{2}
\rightarrow  {\cal D}_{2} + \lambda (\gamma -1) {\cal D}_{1}$ and  $ {\cal
D}_3  \rightarrow {\cal D}_3 + \lambda {\cal D}_2 +
{\lambda^{2}(\gamma -1)\over 2} {\cal D}_{1}$. The compatibility of a
multi-Hamiltonian structure guarantees the complete integrability of a
system of dynamical equations \cite{21}. However, in the case of the polytropic
gas equations, one can check from Eqs. (\ref{2},\ref{4}) that the
conserved  densities
satisfy (this is a particular form of the relations satisfied by the
conserved quantities of a hydrodynamic system)
\begin{equation}
{\partial^{2} H_{n}^{(i)}\over \partial v \partial v} = v^{\gamma -3}
{\partial H_{n}^{(i)}\over \partial u \partial u},\qquad i=1,2 \;{\rm
and}\quad n=0,1,2,\cdots
\end{equation}
Using this, one can prove, in an elegant manner, that the conserved
quantities are in involution with respect to the three Hamiltonian
structures, which proves the complete integrability of both the
systems of equations, namely, the polytropic gas and the elastic
medium. Both can be described by the same Lax function and share the
same conserved quantities \cite{11}.

Let us next use the involution of the conserved quantities to derive
some relations which will be useful within the context of the
supersymmetric polytropic gas. Note that the involution of the charges
with respect to the first Hamiltonian structure implies that
\begin{equation}
\{H_{n}^{(1)},H_{m}^{(1)}\}_{1} = \int dx\,\left({\delta
H_{n}^{(1)}\over \delta u}\partial {\delta H_{m}^{(1)}\over \delta v}
- {\delta H_{n}^{(1)}\over \delta v}\partial {\delta H_{m}^{(1)}\over
\delta u}\right) = 0
\end{equation}
An explicit calculation of the Poisson bracket yields
\begin{equation}
\int dx\,\sum_{s=0}^{[{n\over 2}]+[{m\over 2}]} {u^{m+n-2s-2}v^{s(\gamma
-1)+1}u_{x}\over s(\gamma -1) +1}\,\sum_{l=0}^{s} f(s,l,n,m)
\end{equation}
where
\begin{eqnarray}
f(s,l,n,m) & = &  -(n-2(s-l))(n-2(s-l)-1)(l(\gamma -1)+1)(l(\gamma
 -1))\nonumber\\
 &  &  + (m-2l)(m-2l-1) ((s-l)(\gamma -1)+1)((s-l)(\gamma -1))\label{7}
\end{eqnarray}
The vanishing of the Poisson bracket, then, leads to the relation that
\begin{equation}
\sum_{l=0}^{s} f(s,l,n,m) = 0\label{8}
\end{equation}

Integrable systems can be described by several distinct Lagrangians
\cite{22} 
and it has  been shown recently \cite{20}  that the  polytropic gas  dynamics
allows  four different Lagrangian descriptions. We present the
simplest of these, which will also be useful in our discussion of the
supersymmetric system. Introducing the Clebsh potentials \cite{23}
\begin{equation}
v =  \varphi_x, \qquad  u = \psi_x\label{8'} 
\end{equation} 
we note that the simplest action that describes the polytropic gas
dynamics has the form
\begin{equation}
S = \int dt\,L =\int dt\,dx\,\left (\psi_x\varphi_t - H_{2}^{(1)}
\right)\label{9}  
\end{equation} 
 
Finally, let us note that it is easy to verify that the Miura transformation 
\begin{equation}
v = rp, \qquad u = \frac{1}{(\gamma-1)}\Big ( r^{(\gamma-1)} +
p^{(\gamma-1)} \Big )\label{10} 
\end{equation}
transforms the polytropic gas equations to the equations
\begin{eqnarray} 
\frac{d {~} r}{dt} & = & \frac{1}{(\gamma-1)}\Big (r^{(\gamma-1)} +
p^{(\gamma-1)} \Big )r_x + rp^{(\gamma-2)}p_x \nonumber \\ 
\frac{d {~} p}{dt} & = & \frac{1}{(\gamma-1)}\Big (r^{(\gamma-1)} +
p^{(\gamma-1)} \Big )p_x + pr^{(\gamma-2)}r_x  
 \end{eqnarray}
which can be written as Hamiltonian equations with the first structure
in Eq. (5) and
\[
H = \int dx\, {1\over \gamma (\gamma -1)}\,\left(p r^{\gamma -1} + r
p^{\gamma -1}\right)
\] 
Moreover, this Miura transformation transforms the canonical
Hamiltonian operator ${\cal D}_1$ onto the second Hamiltonian operator
${\cal D}_2$.

\section{$N=1$ Supersymmetric polytropic gas:} 

In this paper, we consider only the non-extended supersymmetrization of
the polytropic gas for arbitrary $\gamma$. As we have already
mentioned in the introduction, we do not have a good understanding of
the Lax description of supersymmetric dispersionless
systems yet \cite{14,15}. Therefore, we have chosen to work directly
at the level of the equations of motion.

From the discussion of the last section and, in particular Eq. (\ref{1}), we
note that we can assign the following dimensions
\begin{equation}
[x] = - 1,\qquad [v] = 2,\qquad [u] = \gamma -1,\qquad [t] = -
\gamma\label{11} 
\end{equation}
In generalizing Eq. (\ref{1}) to the $N=1$ supersymmetric system, we work in
superspace and enlarge the number of dynamical variables to write them
in terms of two fermionic superfields of the forms
\begin{equation}
U(x,\theta) = \eta(x) + \theta u(x),\qquad V(x,\theta) = \xi(x) +
\theta v(x)\label{12}
\end{equation}
where $\eta,\xi$ represent the dynamical fermionic variables, which
are the superpartners of the original bosonic variables, $u,v$,
respectively. Here $\theta$ (with $\theta^{2}=0$) represents the
Grassmann coordinate of the superspace, on which the supercovariant
derivative is defined as 
\begin{equation}
D = {\partial\over \partial\theta} + \theta {\partial\over \partial x}
\end{equation}
which satisfies
\begin{equation}
D^{2} = {\partial\over \partial x}
\end{equation}
This determines the dimension $[\theta] = -{1\over 2}$ so that the
dimensions of the dynamical superfields follow as
\begin{equation}
[U] = \gamma - {3\over 2},\qquad [V] = {3\over 2}\label{13}
\end{equation}

It is clear now that, since $[U_{t}] = 2\gamma - {3\over 2}$ and
$[V_{t}] = \gamma + {3\over 2}$, in generalizing the polytropic gas
equations to the superspace, we can allow for all possible local terms
(involving $U,(DU),U_{x},\cdots$ and $V,(DV),V_{x},\cdots$)
in the dynamical equations on the superspace, which conform to the
appropriate dimensions. Furthermore, we also require that the
equations be such that, in the bosonic limit, they reduce to the
equations for the polytropic gas dynamics. This restricts the
structure of the equations greatly and leads to
\begin{eqnarray}
V_{t} & = & \left(V(DU)\right)_{x}\nonumber\\
U_{t} & = & D\left[{1\over 2} (DU)^{2} + {1\over (\gamma -1)}
(DV)^{\gamma -1} - {(\gamma - 2\kappa)\over \gamma} VV_{x}(DV)^{\gamma
-3}\right]\label{14}
\end{eqnarray}
Here $\kappa$ is an arbitrary parameter. We want to emphasize that
this is a generic feature of supersymmetrizing integrable equations
that one invariably picks up an arbitrary parameter, whose value
becomes fixed (to one or more particular values) when integrability is
imposed \cite{18}. 

Once we have the dynamical equations, with some work, it can be
determined that the system possesses an infinite set of bosonic
conserved charges which have the closed form
\begin{equation}
H_{n}^{(1)} = \int dz\,\tilde{h}_{n}^{(1)} = \int dz\, \sum_{k=0}^{[{n\over
2}]} c(k,n) V (DV)^{k(\gamma - 1)} (DU)^{n-2k}\label{15}
\end{equation}
where $dz = dx\,d\theta$ represents the integration over the
superspace and $c(k,n)$'s denote the constants defined in
Eq. (\ref{3}). These charges reduce in the bosonic limit to the
\lq\lq Eulerian'' conserved charges in Eq. (\ref{2}) and have the
following  explicit forms for the first few.
\begin{eqnarray}
H_{0}^{(1)} & = & \int dz\, \tilde{h}_{0}^{(1)} = \int dz\,
V\nonumber\\
H_{1}^{(1)} & = & \int dz\,\tilde{h}_{1}^{(1)} = \int dz\,
V(DU)\nonumber\\
H_{2}^{(1)} & = & \int dz\, \tilde{h}_{2}^{(1)} = \int
dz\,V\left[{1\over 2} (DU)^{2} + {(DV)^{\gamma -1}\over \gamma (\gamma
-1)}\right]\nonumber\\
H_{3}^{(1)} & = & \int dz\, \tilde{h}_{3}^{(1)} = \int dz\,
V\left[{1\over 3!} (DU)^{3} + {1\over \gamma (\gamma -1)} (DV)^{\gamma
-1} (DU)\right]
\end{eqnarray}

These \lq\lq Eulerian'' charges can be easily checked to be conserved
for any value of $\kappa$. In fact, under the evolution of Eq. (\ref{14}), it
is straightforward to show that
\begin{eqnarray}
{dH_{n}^{(1)}\over dt} & = & \int dz\,\left[ -\sum_{k=1}^{[{n\over
2}]} c(k,n) {k(\gamma -1)(k(\gamma -1)+1)\over n-2k+1}
V(DV_{x})(DV)^{k(\gamma -1)} (DU)^{n-2k+1}\right.\nonumber\\
 &  & \quad\left. + \sum_{k=0}^{[{n\over 2}]-1} c(k,n) (n-2k)
V(DV_{x})(DV)^{(k+1)(\gamma -1) -1} (DU)^{n-2k-1}\right]\nonumber\\
 & = & \int dz\, \sum_{k=1}^{[{n\over 2}]} \left[- {k(\gamma -1)
(k(\gamma -1)+1)\over n-2k+1} c(k,n) + (n-2k+2)
c(k-1,n)\right]\nonumber\\
 &  & \qquad \times V(DV_{x}) (DV)^{k(\gamma -1)} (DU)^{n-2k+1}\nonumber\\
 & = & 0\label{16}
\end{eqnarray}
where we have used the fact that the quantity in the square bracket
vanishes because of the structure of $c(k,n)$ (see Eq. (\ref{3})).

Once we have the conserved quantities, we can think of them as the
Hamiltonians and look for a Hamiltonian description of the system. It
turns out that the system of equations in Eq. (\ref{14}) has a Hamiltonian
description only for $\kappa =1$. Namely, with the Hamiltonian
structure
\begin{equation}
{\cal D}_{1} = \left(\begin{array}{cc}
0 & D\\
D & 0
\end{array}\right)\label{17}
\end{equation}
the $N=1$ supersymmetric polytropic gas equations can be written in
the Hamltonian form
\begin{equation}
\left(\begin{array}{c}
U_{t}\\
V_{t}
\end{array}\right) = {\cal D}_{1} \left(\begin{array}{c}
{\delta H_{2}^{(1)}\over \delta U}\\
{\delta H_{2}^{(1)}\over \delta V}
\end{array}\right)
\end{equation}
only if $\kappa =1$. This, therefore, selects out the particular value
of the arbitrary parameter $\kappa$. However, integrability is yet to
be shown, even though the presence of an infinite number of conserved
quantities is suggestive.

We would choose $\kappa =1$ from now on so that the $N=1$
supersymmetric polytropic gas equations have the form
\begin{eqnarray}
V_{t} & = & \left(V(DU)\right)_{x}\nonumber\\
U_{t} & = & D\left[{1\over 2} (DU)^{2} + {1\over (\gamma -1)}
(DV)^{\gamma -1} - {(\gamma -2)\over \gamma} VV_{x} (DV)^{\gamma
-3}\right]\label{18}
\end{eqnarray}
Before proving the complete integrability of this system, let us bring
out some essential features of this system. First, in components, the
equations take the form
\begin{eqnarray}
\xi_{t} & = & \left(\xi u\right)_{x}\nonumber\\
v_{t}  & = & \left(uv - \xi \eta\right)_{x}\nonumber\\
\eta_{t} & = & \eta_{x} u + {2\over \gamma} \,\xi_{x} v^{(\gamma - 2)}
+ {(\gamma -2)\over \gamma} \,\xi v_{x}v^{(\gamma -3)}\nonumber\\
u_{t} & = & uu_{x} + v^{(\gamma -2)} v_{x} - {2(\gamma -2)\over \gamma}
\xi\xi_{xx} v^{(\gamma -3)} - {(\gamma - 2)(\gamma -3)\over
\gamma}\,\xi\xi_{x} v_{x} v^{(\gamma -4)}
\end{eqnarray}
As is clear, this is a system of coupled equations unlike the SUSY -~B
extension that we will describe in the next section.  

Second, we note that for $\gamma =2$, our system of equations in (\ref{18})
reduces to the dispersionless limit of the sTB equations
\cite{15}. For  $\gamma
=3$, defining two new superfields as before, $A = U+V$ and $B = U-V$,
we obtain,
\begin{eqnarray}
A_{t} & = & D\left[ {1\over 2} (DA)^{2} - {1\over 3}(A - B)A_{x} -
{1\over 6} (A - B) B_{x}\right]\nonumber\\
B_{t} & = & D \left[ {1\over 2} (DB)^{2} + {1\over 6} (A - B) A_{x} +
{1\over 3} (A - B) B_{x}\right]
\end{eqnarray}
We see that, unlike the bosonic equations in (\ref{1'}), these are coupled
equations, which clearly decouple in the bosonic limit.

Unlike the bosonic system, where there are three distinct Hamiltonian
structures, the supersymmetric equations in (\ref{18}) do not seem to allow
any other Hamiltonian structure (at least, we have not succeeded in
finding them). It can be checked that the structure
\begin{equation}
{\cal D}_{2} = \left(\begin{array}{cc}
\begin{array}{c}
D^{2}V(DV)^{\gamma -3} + V(DV)^{\gamma -3}\\
DV(DV)^{\gamma -3}D
\end{array} & \begin{array}{c}
D^{2}U + (\gamma -2)UD^{2}\\
(\gamma -1) DUD
\end{array}\\
\begin{array}{c}
(\gamma -2) D^{2}U + UD^{2}\\
(\gamma -1) DUD
\end{array} & D^{2}V + VD^{2} - (\gamma -2)DVD
\end{array}\right)\label{19}
\end{equation}
gives the first three flows of the supersymmetric polytropic gas
hierarchy. However, it fails at the next higher order and it can be
checked that, although ${\cal D}_{2}$ has the right symmetry
properties, it fails to satisfy the Jacobi identity. We have analyzed
various deformations of this structure and it seems to us that this
system of equations does not admit any local higher order Hamiltonian
structure. 

Since the system of equations does not allow a second Hamiltonian
structure, the usual proof of involution of charges does not
automatically hold. Nonetheless, let us note from the structure of the
conserved charges in Eq. (\ref{15}) that they satisfy identities analogous to
the bosonic case, namely,
\begin{eqnarray}
{\partial^{2} \tilde{h}_{n}^{(1)}\over \partial (DV)\partial (DV)} & =
& (DV)^{\gamma -3} {\partial^{2} \tilde{h}_{n}^{1)}\over \partial
(DU)\partial (DU)}\nonumber\\
{\partial^{3} \tilde{h}_{n}^{(1)}\over \partial V\partial (DV)\partial
(DV)} & = & (DV)^{\gamma -3} {\partial^{3} \tilde{h}_{n}^{(1)}\over
\partial V\partial (DU)\partial (DU)}
\end{eqnarray}
However, in spite of such relations, we have not managed to find an
elegant proof of involution of charges as in the bosonic
case. Instead,  we can show by brute force that the charges are in involution
with respect to the Hamiltonian structure ${\cal D}_{1}$, namely, it
can be shown by direct calculations that
\begin{eqnarray}
\left\{H_{n}^{(1)},H_{m}^{(1)}\right\} & = & \int dz\,
\sum_{s=0}^{[{n\over 2}]+[{m\over 2}]} {\left(V(DU_{x})(DV)^{s(\gamma
-1)} + V_{x}VU_{x} (DV)^{s(\gamma -1)-1}\right) (DU)^{m+n-2s-2}\over
s(\gamma -1) + 1}\nonumber\\
 &  & \qquad \times \sum_{l=0}^{s} f(s,l,n,m)\nonumber\\
 & = & 0\label{20}
\end{eqnarray}
where $f(s,l,n,m)$ is defined in Eq. (\ref{7}) and we have used the identity
in Eq. (\ref{8}) in the final step. This shows that the infinite set of
charges are  in
involution thereby proving the complete integarbility of the system.

As we have seen, in the bosonic case, the polytropic gas equations
possess two infinite sets of conserved charges. Although only one set
of infinite conserved charges is enough to prove integrability, in the
case of the polytropic gas, the existence of a second set is an
accident, which is understood because of the peculiar feature that
both the polytropic gas and the elastic medium equations have the same
Lax description and share the same conserved charges \cite{11}. It is worth
inquiring about the second infinite set of conserved charges in the
case of supersymmetric polytropic gas equation. Unfortunately, we have
found only one other charge, namely,
\begin{equation}
H_{0}^{(2)} = \int dz\, \tilde{h}_{0}^{(2)} = \int dz\,U\label{20'}
\end{equation}
which is conserved under the evolution of the supersymmetric
polytropic gas system. 
This, of course, reduces, in the bosonic limit, to the bosonic
charges $H_{0}^{(2)}$ in Eq. (\ref{4}). 

One can, of course, ask what is the generalization of the elastic
medium equations to the superspace and whether they have any relation to
the supersymmetric polytropic gas, as is the case in the bosonic
limit. Without going into details, let us note here that the general
one  parameter family of supersymmetric extension of
the elastic medium equations turns out to be
\begin{eqnarray}
V_{t} & = & U_{x}\nonumber\\
U_{t} & = & {1\over \gamma -1} D\left[\left(1-\kappa + {1\over \gamma
-2}\right) (DV)^{\gamma -2} + \kappa D\left(V(DV)^{\gamma
-3}\right)\right]
\end{eqnarray}
This has the right bosonic limit for any value of the arbitrary
parameter $\kappa$. Furthermore, with the
supersymmetric Hamiltonian structure in Eq. (\ref{17}), it is simple to check
that, this equation is Hamiltonian only for $\kappa =1$, where the
Hamiltonian is given by
\begin{equation}
\tilde{H} = \int dz\, \left[{1\over 2} U (DU) + {1\over (\gamma
-1)(\gamma -2)} V (DV)^{\gamma -2}\right]
\end{equation}
The dynamical equations, in this case, take the form
\begin{eqnarray}
V_{t} & = & U_{x}\nonumber\\
U_{t} & = & D\left[{1\over \gamma -2} (DV)^{\gamma -2} - {(\gamma
-3)\over (\gamma -1)} VV_{x}(DV)^{\gamma -4}\right]
\end{eqnarray}
We can think of this to be the $N=1$ supersymmetric generalization of
the elastic medium equation. It is manifestly supersymmetric and
reduces to the elastic medium equations in the bosonic limit and is
Hamiltonian. In
components, the system of equations has the form
\begin{eqnarray}
\xi_t & = & \eta_x \ \\
v_t & = & u_t \ \\
\eta_t & = & {1\over \gamma -1}\Big ( 2\xi_xv^{(\gamma-3)} + (\gamma-3)\xi v_x
v^{(\gamma-4)}  -(\gamma-3)\xi \xi_{xx} v^{(\gamma-4)} \Big ) \ \\
u_t & = & {1\over \gamma -1}\Big ( (\gamma-1)v_xv^{(\gamma-3)} -
(\gamma-3) (\gamma-4)
\xi\xi_xv_x v^{(\gamma-5)} \Big ) 
\end{eqnarray}
It is
interesting to note that, in addition to the conserved charge in
Eq. (\ref{20'}) as well as $\tilde{H}$, the charges
$H_{0}^{(1)},H_{1}^{(1)}$ of
Eq. (\ref{15}) are  also
conserved under this flow. But, the infinite set of charges in Eq. (\ref{15})
are clearly not conserved under this flow. Furthermore, we have not
found any more conserved charges associated with this
system. Therefore, it is not at all clear whether the supersymmetric
elastic medium equations are even integrable. This remains an open
question.

We note here that the supersymmetric polytropic gas equations admit a
Lagrangian description. Following the discussion in the bosonic case,
we define a supersymmetric analog of Clebsch potentials as (see
Eq. (\ref{8'})) 
\begin{equation}
V = \Phi_{x} = \left(\alpha + \theta \varphi\right)_{x},\qquad U =
\Psi_{x} = \left(\beta + \theta \psi\right)_{x}\label{21}
\end{equation}
where $\alpha,\beta$ are the fermionic fields which are the
superpartners of the bosonic fields $\varphi,\psi$ respectively. In
terms of these potentials, the action for the supersymmetric
polytropic gas has the form
\begin{equation}
S = \int dt\,L = \int dt dz\,\left(\Phi_{x}(D\Psi_{t})  -
H_{2}^{(1)}\right)\label{22}
\end{equation}
which, in components has the explicit form
\begin{equation}
S = \int dt dx\,\left(\varphi_{x}\psi_{t} -\alpha_{x}\beta_{tx} -
\varphi_{x}\psi_{x}^{3} - {2\over \gamma (\gamma -1)}
\varphi_{x}^{\gamma} + 2\alpha_{x}\beta_{xx}\psi_{x} + {2\over \gamma}
\alpha_{x}\alpha_{xx}\varphi_{x}^{\gamma -2}\right)
\end{equation}
The same set of Clebsch potentials also allows a Lagrangian
description of the supersymmetric elastic medium equation through the
action
\begin{equation}
S = \int dt dz\,\left(\Phi_{x} (D\Psi_{t}) - H_{1}^{(2)}\right)
\end{equation}
 
Let us comment now on why our supersymmetric polytropic
gas system has only one local Hamiltonian structure 
and possesses only the supersymmetric \lq\lq Eulerian'' series of
local  conserved
charges (and not a second infinite set of conserved charges).  First,
note  that the bosonic  polytropic gas
equation (\ref{1}), for $\gamma=4$, can be thought of as the dispersionless
limit of the Boussinesq equation. 
On the other hand, the second Hamiltonian structure of this equation
(see Eq. (\ref{5})), corresponds to the  $W_3$ algebra.  
It is well known that there is no nontrivial $N=1$ supersymmetric extension
of the Boussinesq equation \cite{24} as well as the  $W_3$ algebra, except for
the SUSY B-extension, which we will discuss in the next section. From
the nonexistence of supersymmetric extension 
of $W_3$  algebra alone, we cannot, of course, conclude that there are no
possible supersymmetrizations in the dispersionless limit. However, 
using computer and symbolic computations, we have checked that it
is impossible to find such local structures as well as the
supersymmetric analog of the Miura transformation, Eq. (\ref{10}), in the
dispersionless limit. The non-existence of the second Hamiltonian
structure, for the supersymmetric polytropic gas, also implies that
there does not exist a recursion operator that is factorizable and can
be written in terms of two Hamiltonian structures. This, however, does
not affect integrability of the system, as we have explicitly
demonstrated that the $N=1$ supersymmetric polytropic gas system is
completely integrable.

\section{SUSY -B extension of the polytropic gas dynamics:}

The supersymmetric -B extensions \cite{19} are simple supersymmetrizations of a
bosonic integrable model that are automatically integrable. The basic
idea is to replace a classical bosonic variable $u\rightarrow (DU)$ in
the bosonic equation \cite{19,25}. Thus, in the case of the polytropic gas
equation, for example, if we let ($U,V$ are fermionic superfields,
whose components can be taken as in Eq. (\ref{12}))
\begin{equation}
u\rightarrow (DU),\qquad v\rightarrow (DV)\label{23}
\end{equation}
then, the dynamical equations would lead to
\begin{eqnarray}
V_{t} & = & D\left((DV)(DU)\right)\nonumber\\
U_{t} & = & D\left[{1\over 2} (DU)^{2} + {(DV)^{\gamma -1}\over
(\gamma -1)}\right]\label{24}
\end{eqnarray}
which would correspond to the $N=1$ SUSY -B extension of the
polytropic gas equation. These equations have the following explicit
forms in components
\begin{eqnarray}
\eta_t & = & \eta_xu+v\xi_x, \nonumber\\
v_t & = & \left ( vu \right )_x, \nonumber\\
\xi_t & = & \xi_xu+\eta_xv^{(\gamma-2)}, \nonumber\\
u_t & = & uu_x + vv^{(\gamma-2)}\label{25}
\end{eqnarray}
As is clear, the bosonic equations are completely unchanged by such a
supersymmetrization, namely, there is no  interaction terms with the
fermionic variables. This is similar, in that respect, to the set of
supersymmetric Chaplygin gas equations obtained in \cite{17}. However, these
equations are not identical to those in \cite{17}. In fact, note that the -B
extension, in Eq. (\ref{24}), naturally leads to $N=1$ supersymmetry
and,  consequently,
there are two fermionic variables, $\eta,\xi$ here, which are the
superpartners of the two bosonic variables $v,u$ respectively, whereas
the model in \cite{17} has only one fermionic variable. This is easily
understood from the fact alluded to in the introduction, namely, the
system of equations in \cite{17} are obtained from a starting superstring
theory, which has a local {\em kappa} supersymmetry that needs to be
fixed leaving one with only a single fermionic variable.

There is another important difference between the two systems. The
Hamiltonian structures for SUSY -B extensions are odd. It is easy to
see that the Hamiltonian structures for the SUSY -B extension of any
integrable model is given by (acting on a delta function in
superspace) 
\begin{equation}
{\cal D}_n^{({\rm B})} = D^{-1} {\cal D}_n^{({\rm bosonic})}
D^{-1}\label{26} 
\end{equation}
where $D^{-1} = \partial^{-1}D$ and ${\cal D}_{n}^{({\rm bosonic})}$
represents the Hamiltonian structure for the bosonic integrable
system.  Using for ${\cal D}_{n}^{({\rm bosonic})}$, the three distinct
Hamiltonian structures in Eq. (\ref{5}), we see that SUSY -B extension of
the polytropic gas dynamics also can be described by three distinct
Hamiltonian structures, which, however, are odd. The corresponding
Hamiltonians can also be easily constructed and which are Grassmann
odd (This is the peculiar feature of SUSY -B extensions). In contrast,
both the Hamiltonian and the Hamiltonian structure in \cite{17} are of the
usual kind (even). Presumably, one can think of extensions, such as in
\cite{17} as the $N={1\over 2}$ SUSY -B extension of an integrable
system. However, we have not investigated this question in more detail.

\section{Conclusion:}

In this letter, we have constructed the $N=1$ supersymmetric extension
of the polytropic gas dynamics as well its SUSY -B extension. We have
shown that the supersymmetric polytropic gas equations are Hamiltonian
only for a specific value of an arbitrary parameter. We have
constructed the infinite set of conserved charges, the \lq\lq
Eulerian'' charges, and have shown that they are in involution,
thereby proving the complete integrability of the system. Unlike the
bosonic system, in the supersymmetric system, the second infinite set
of conserved charges - the \lq\lq Lagrangian'' ones - do not exist. In
stead, we have found only one such charge, in addition to the \lq\lq
Eulerian'' ones. A Lagrangian description for the system, through the
Clebsch superpotentials, has been constructed. We have tried to argue
why a second Hamiltonian structure for such a system does not appear
to exist. We have also
constructed the supersymmetric elastic medium equations, which,
however, do not appear to be integrable. We have commented on the
similarities and differences between the SUSY -B extension of the
polytropic gas and an earlier supersymmetrization of this system. We
have pointed out several open issues associated with this system. 

\section*{Acknowledgments} One of the author (ZP) would like to thank
Dr M.Pavlov  for  fruitful discussions on the classical polytropic
gas dynamics. This work was supported in part by US DOE grant
no. DE-FG-02-91ER40685 as well as by NSF-INT-0089589.

\end{document}